\documentclass{knawproc}

\input{psfig.tex}
\begin{document}

\begin{opening}

\title{X-Ray Observations of Radio Galaxies}

\author{D. E. Harris}
\addresses{%
  Smithsonian Astrophysical Observatory, 60 Garden St., Cambridge, MA
02138 USA\\
}

\end{opening}


\begin{abstract}
We review some of the ways that X-ray observations provide unique
information on radio galaxies.  Thermal bremsstrahlung X-ray emission
provides detailed data on ambient densities and temperatures.  These
parameters in turn can be used for pressure balance calculations and
can demonstrate how the ambient gas affects radio source structure.
Additionally, many signatures of the interaction of radio jets and
lobes with the hot gas are found in high resolution X-ray maps.

Non-thermal X-ray emission from knots and hotspots of radio jets can
give us constraints on the relativistic electron population for
energies greater that that normally sampled in the radio (in the case
of synchrotron emission) or can give us an independent estimate of the
average magnetic field strength (if inverse Compton emission is the
origin of the X-rays).  From recent ROSAT HRI observations of 3C 390.3
and 3C 120, we show evidence that X-ray emission from knots and
hotspots appears to be associated with regions of large gradients in
the radio surface brightness; i.e. at the location of powerful shocks.
\end{abstract}


\section{Introduction}

X-ray observations of radio galaxies provide vital data unobtainable
at other wavelengths.  Most of the constraints derived from X-ray
observations can be divided into one of two areas: a study of the hot
ambient gas from thermal bremsstrahlung emission or investigation of
non-thermal emission from particular radio features which in turn
constrains the parameters which describe the magnetic field and
relativistic electrons responsible for the radio emission.  We review
here several topics in each area, provide some new data from ROSAT
High Resolution Imager (HRI) observations, and append a short section
on the situation at high z.

\section{Thermal Emission from Hot Gas}

Hot, X-ray emitting gas appears to be a ubiquitous feature of all
reasonably deep potential wells associated with bulge systems.
Although the bulges in spirals are often not detectable with current
satellites, this is not the case for normal or massive ellipticals
(e.g. Davis and White III 1996).  These systems are characterized by
gas with temperatures of a few keV (i.e. a few $\times~10^{7}$~K) and
the X-ray satellites EINSTEIN, ROSAT, and ASCA have been well matched
to their primary emitting band.

Generally, the X-ray brightness distribution is well described by a
modified King distribution (a.k.a. `Beta Model'):

$$I(r)\propto an_{o}^{2}\left[1+\left({r\over
a}\right)^{2}\right]^{-(3\beta-{1\over 2})}$$

$$n(r)\propto n_{o} \left[1+\left({r\over
a}\right)^{2}\right]^{-{3\beta \over 2}}$$

\noindent
where I(r) is the X-ray surface brightness as a function of radius, r;
a is the core radius; n(r) is the electron density as a function of r;
and n$_{o}$ is the electron density at r=0.

Since the observable extent is tens of kpc (hundreds of kpc for
cluster atmospheres), the X-ray observations give an estimate of the
electron density over a region which often encompasses the radio source.
At the same time, an estimate of the temperature is obtained from
those detectors which have reasonable spectral resolution.

\subsection{Pressure Balance}

A potentially powerful diagnostic is provided by the expectation that
some radio features will be in pressure balance with the external
medium.  In practice, there are several problems which limit the
usefulness of the concept.  

Do most radio features exclude ambient gas?  If not, we must include a
thermal contribution to the internal pressure, and this is normally
difficult to estimate.  For the case of the lobes of Cygnus A, Carilli
et al. (1994) and Clarke et al. (1997) argue that the X-ray
observations show that the lobes exclude the hot cluster gas.

In almost all radio structures, we don't know the internal
(non-thermal) pressure, so we compute the minimum pressure.  This
assumes that only the observed synchrotron spectrum is used to
estimate the relativistic electron population (i.e. no contribution
from electrons which radiate below the lowest observed frequency);
equipartition holds between the magnetic field and particle
energy densities; the filling factor is unity; and there is no
significant contribution to the energy density from relativistic
protons (see e.g. Harris et al. 1995 for further discussions).

If the radio feature is moving with respect to the ambient medium, a
ram pressure component must be added to the pressure balance equation.
If the internal pressure is greater than the external pressure, then
we expect expansion.  Since it is difficult to estimate the magnitude
of the ram pressure, it is preferable to use radio features whose
boundaries are thought to be stationary or moving very slowly.

In many cases there is the added complication of the projection
factor.  For sources such as Cygnus A which is most likely close to
the center of the observed hot gas distribution we can reliably obtain
a thermal pressure at the positions of the hotspots.  However, for
other sources such as tailed radio galaxies (TRG) in clusters, we
don't know how far out in the cluster atmosphere the source really is,
and thus we have only an upper limit on the external gas pressure.

The most interesting application would be a detailed study of radio
features for which all of the above uncertainties were minimized.  Our
expectation would be that the external gas pressure should be greater
than the minimum non-thermal pressure, and the magnitude of the
difference would provide an estimate of the contribution of protons
and/or low energy electrons to the internal pressure.  Such a study
was made by Feretti et al. (1990) for radio galaxies at the centers of
rich clusters.  They found thermal pressures always greater than
minimum non-thermal pressures and from the magnitude of the pressure
ratios, derived ranges of the filling factor (0.03 to 0.3) or of the
ratio of energy in protons to that in electrons (6 to 75) which would
restore pressure balance.

\subsection{Buoyancy}

Since we have evidence that most radio features exclude the ambient
gas and thus their internal pressure is dominated by non-thermal
components, it follows that these will be lighter than the external
medium and experience buoyancy forces.  We can expect that these
forces will be manifest whenever the external pressure gradient is
large enough to produce a significant (compared to other fluctuations
of the pressure) difference in force between one part of the radio
feature and another.  The two conditions of interest are large
gradients in pressure and/or large sources (e.g. TRGs in cluster
atmospheres).  Note also that if the radio source is expanding
supersonically, buoyancy is irrelevant.

For the most part, buoyancy has been suggested as a plausible
explanation of changes of position angle for lower brightness regions.
Examples are 26W20 (Harris, Costain, and Dewdney 1984); a TRG in Abell
115 (Gregorini and Bondi, 1989); and the TRG in the 0335+09 cluster (
Sarazin, Baum and O'Dea 1995).  Worral, Birkinshaw, and Cameron (1995)
present a more quantitative analysis for the buoyancy forces
operating on NGC 326.

\subsection{Hydrodynamics of Interactions}

Many of the hypothesized attributes of radio galaxies become
accessible via X-ray investigations of the hot ambient gas.  From
ROSAT HRI observations of Cygnus A (Carilli, Perley, and Harris 1994)
we found evidence for cavities in the ambient gas caused by the radio
lobes.  We also suggested that symmetric features of enhanced X-ray
brightness were caused by the longer path length through the sheath of
compressed gas between the (hypothetical) bow shock and the radio
lobe.  These results which relied on approximate analytical
calculations were corroborated by hydrodynamical simulations (Clarke,
Harris, and Carilli 1997) which also demonstrated the vastly different
morphologies expected for the X-ray surface brightness of a modified
King distribution of hot gas disturbed by a powerful radio galaxy.  In
particular, the match or mis-match of the X-ray spectral sensitivity
compared to the temperature of the ambient medium has a strong effect
on the resulting X-ray map.  For Cygnus A, with a temperature close to
4 keV, the ROSAT HRI is well matched to the ambient temperature, but
not to the gas at the leading edge of the bow shock, which is expected
to be much hotter.  Hence we found the cavities and the sheath well
away from the leading edge.  The exciting prospects for AXAF data are
the detection of the leading bow shock at the higher energies and
more spatial detail and spectral resolution for the known features
at lower energies.

\subsection{Local Weather}

By `weather' we mean `Which way is the wind blowing?'.  The problem of
winds in the ICM has only recently received serious attention because
the idea that many clusters show attributes of recent or ongoing
mergers now creates a natural explanation for large scale gas motions
with respect to individual galaxies.  A long standing problem has been
the explanation for the morphology of wide angle tailed (WAT) radio
galaxies.  Since WATs are normally associated with the dominant
cluster galaxy, there is good reason to believe that their velocity
with respect to the cluster potential is nil (Eilek et al. 1978).
However, as shown by numerical simulations (Loken, Roettiger, and
Burns 1995; Roettiger, Burns, and Loken 1996) the winds produced by
mergers are capable of generating the WAT morphology.

A related, perhaps more common phenomenon is the bending of radio jets
which often occurs as the jet leaves the ISM and experiences the
ICM. Gregorini and Bondi (1989) invoke this explanation for the TRG in
Abell 115.

\begin{figure}
\centerline{
\psfig{figure=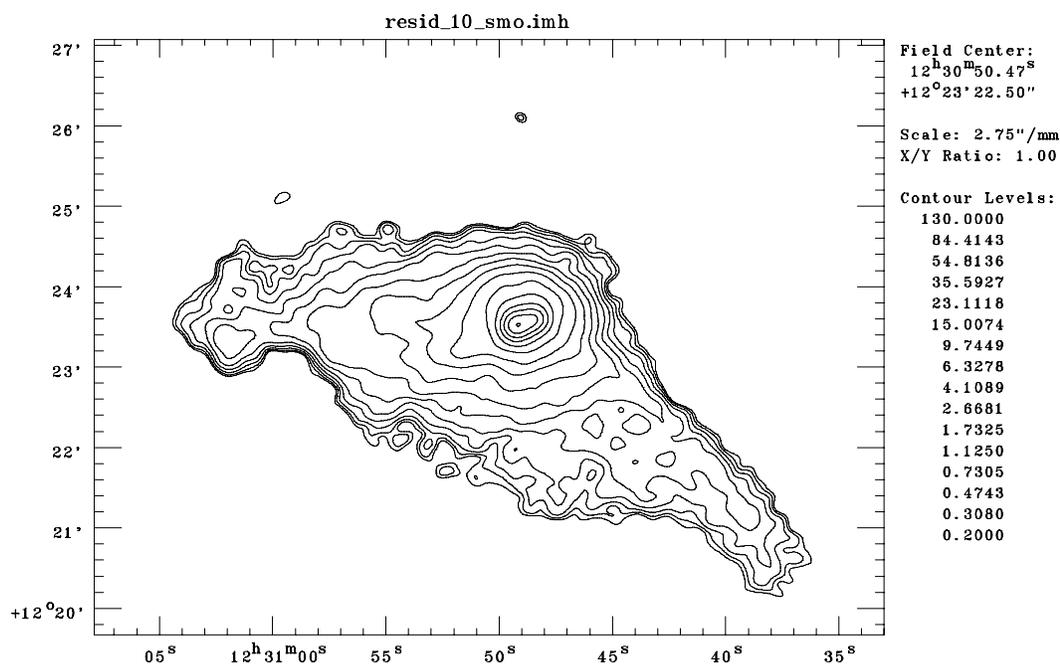,width=15cm}
}
\caption{\label{fig1} A ROSAT HRI map of M87 from Harris, Biretta and
	 Junor (1998).  A power law model has been subtracted to leave
	 only the asymmetrical structures, but we did not attempt to
	 fit the inner section.  A Gaussian smoothing function of FWHM
	 = 10'' was applied.  Contour levels are logarithmic, starting
	 at 0.2, and ending in 130 counts per 1.0'' pixel.  NB: the
	 zero level is arbitrary; the subtraction produced negative
	 areas at large radii. }
\end{figure}

A striking example of a possible interface between the ISM of M87 and
a moving ICM in the Virgo cluster is afforded by recent ROSAT HRI
observations (Harris, Biretta, and Junor 1998).  The X-ray `spur' (see
Figure~\ref{fig1}) which extends about 4.5 arcmin to the SW is most
likely thermal emission since it does not correspond to any radio
feature.  Together with the X-ray arm which extends to the East, one
obtains the distinct impression of a bow shock (with large gradient on
the leading edge) whose principal axis would be in PA $\approx$ 150
degrees.  The idea of a merger in the Virgo cluster is suggested also
by Binggeli (1998) who presents substantial evidence that the
so-called `M86 subclump' is merging with the M87 subclump.  If these
ideas are correct, the wind model might be useful in understanding the
large scale radio structure of M87 which has long been thought of as a
WAT in projection (i.e. the major extent of the radio arms are along
the line of sight).

\subsection{Effects of Pressure Gradients}

When a radio jet crosses a region where the external pressure is
falling sharply, the jet can lose its tight collimation.  Several
examples of this behavior have been noted, including several of the
sources cited in the previous section (e.g. when a jet passes from the
ISM to the ICM).  Another effect could be the genesis of an internal
shock.  If one traces back the boundary of the SW spur in
figure~\ref{fig1}, it passes close to the position of knot A in the
M87 jet.  This raises the possibility that the occurrence of this
strong shock is caused by a sudden change in the external pressure
(see e.g. Hooda and Wiita 1998).

\section{Non-thermal Emission: Hotspots and Knots in Radio Jets}

In addition to the wealth of information afforded by studies of
thermal X-ray emission around radio galaxies, the non-thermal
emissions detected from a handful of radio features provides
constraints on acceleration processes, the characteristics of the
population of relativistic electrons responsible for the radio
emission, and the magnetic field strength.  The two primary emission
mechanisms normally considered for these features are inverse Compton
(IC) emission and synchrotron emission.  The IC process can work on
any photon distribution, but here we are mostly concerned with either
IC from the 3k background photons (IC/3k) or synchrotron self Compton
(SSC) emission.  IC/3k is expected from radio features of low surface
brightness (i.e. weak magnetic fields, Harris and Grindlay, 1979)
whereas the SSC process requires compact, high brightness radio
structures in order to be detectable by current X-ray systems.  It is,
however, important to remember that both types of IC emission are
mandatory physical processes which occur in all (radio) synchrotron
sources.

The problem of synchrotron X-ray emission revolves around the poorly
understood questions of the acceleration mechanisms.  If we invoke the
classical shock model, then we need to devise conditions which allow
the resulting population of relativistic electrons to extend to
Lorentz energy factors, $\gamma~\approx~10^{7}$ in order to explain
X-ray generation in typical field strengths of order 100 $\mu$G.

\subsection{IC/3k Emission}

Because the major contribution to the photon energy density occurs at
a frequency $\approx~1.6\times~10^{11}$~(1+z) Hz, the electrons
responsible for a given X-ray energy will have the same Lorentz factor
regardless of the redshift (e.g. for 1 keV photons, the relevant
electrons have $\gamma$ = 1069 (Harris and Grindlay, 1979).  Since the
magnitude of the 3k energy density is known, any X-ray detection of
IC/3k emission can be used to obtain the amplitude of the electron
spectrum at the relevant energy.  If one is prepared to make the
extrapolation of the electron spectrum from the observed X-ray point
to that region defined by the radio emission, we can thus derive a
measure of the average magnetic field strength.  There have been many
unsuccessful attempts to do this (e.g. Harris et al. 1995).

There are however two convincing cases: the radio lobes of Fornax A
and a relic radio source associated with the cluster Abell 85.  Using
the ROSAT PSPC, Feigelson et al. (1995) argue that the observed X-rays
represent the detection of IC/3k emission based primarily on spatial
analysis.  Using ASCA, Kaneda et al. (1995) demonstrate a consistent
result based on a spectral analysis.  In both investigations, a field
strength of a few $\mu$G is derived, roughly consistent with
expectations from equipartition arguments.  Recently, Bagchi, Pislar,
and Lima Neto (1998) have combined low frequency radio and PSPC X-ray
data.  They find a magnetic field strength of 0.95 $\mu$G for a steep
spectrum radio source which lies at a projected distance of
$\approx$~700 kpc from the center of Abell 85
(H$_{\circ}$=50~$km~s^{-1}~Mpc^{-1}$).

Hwang (1997) has suggested that an observed excess of extreme ultra
violet (EUV) emission from the Coma and Virgo clusters may be caused
by IC/3k emission and Sarazin and Lieu (1998) have extended this idea
to other clusters, even those without observable radio halos.  One of
the problems of these models is that $\gamma\approx$~300 in order to
produce the observed EUV emission, so that the extrapolation of the
electron spectrum to/from the observed radio regime involves greater
uncertainties.

\subsection{SSC Emission}

Except for (unresolved) emission from the cores of radio galaxies and
quasars, SSC emission has been reliably established only in the radio
hotspots of Cygnus A (Harris, Carilli, and Perley, 1994).  By
`reliably' we mean that the radio structure has been resolved and the
radio spectrum is well defined so that a good estimate of the
synchrotron photon energy density can be calculated.  Furthermore, the
resulting values of the average magnetic field strength
(158$\pm$17~$\mu$G for hotspot A, and 246$\pm$22~$\mu$G for hotspot D)
are in good agreement with the conventional equipartition values: 134
to 183~$\mu$G for hotspot A and 192 to 262~$\mu$G for hotspot D.

Although the SSC model for the hotspots is completely satisfactory,
another possible model is synchrotron emission associated with the so
called ``proton induced cascade'' (PIC) process (Mannheim, Krulls, and
Biermann 1991).  The PIC is a sister process to SSC in that it
requires high photon energy densities, but relies on extremely high
energy protons ($\gamma~\approx~10^{11}$).  Instead of a steep power
law of relativistic electrons covering many decades in energy, the
electrons responsible for the observed X-ray emission are supplied via
pair production form higher energy photons.  If PIC emission were
responsible for the X-ray emission from the hotspots of Cygnus A, the
resulting magnetic field strength would have to be greater than
500$\mu$G compared to the value of 100 to 200 $\mu$G derived from
SSC. Thus the lower field values can be thought of as strict lower
limits to the average field strength, regardless of the emission
process responsible for the X-rays.

One of the implications of the SSC model concerns the `fluid' of the
jets.  If the jets were to consist of normal matter (electrons and
protons), we have every reason to believe that the terminal shock of
the jet (the radio hotspot) would accelerate the protons as well as
the electrons and that the proton contribution would dominate the
relativistic particle energy density (i.e. the (1+k) factor would be
greater than 50).  However, the equipartition field with which the SSC
field agrees is calculated on the basis of little or no energy density
from protons.  If equipartition is valid, this implies that there are
no protons in the jet which would then be required to contain
electrons and positrons.  On the other hand, if PIC is the primary
emission process, the magnetic energy density is much greater and
equipartition is maintained with a dominating contribution to the
particle energy density from protons.  Unfortunately we have been
unable to devise a definitive test to differentiate between SSC and
PIC emission for the Cygnus A hotspots.  The gamma ray satellites
currently operational are not sensitive enough to detect the predicted
higher energy PIC emission.

\subsection{Synchrotron Emission}

\begin{figure}
\centerline{
\psfig{figure=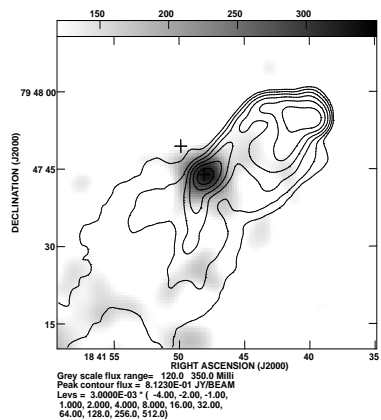,width=5cm,height=6cm}
}
\caption{\label{fig2} The northern hotspot(s) in 3C 390.3 The grey
scale is a 5$^{\prime\prime}$ smoothed X-ray map and the contours show
the radio brightness.  The contours are logarithmic (factors of two)
with the first level at 3 mJy/beam.  The cross to the NE of hotspot B
shows the approximate position of the adjacent dwarf galaxy. }
\end{figure}

Synchrotron emission has been the process `of choice' for non-thermal
models of X-ray emission from knots and hotspots associated with radio
jets.  A primary example is knot A in the M87 jet (Biretta, Stern, and
Harris 1991).  The optical morphology mimics the radio structure and
the optical emission is polarized.  An extrapolation of the optical
spectrum of knot A to the X-ray flux density is consistent with a
single power law (albeit steeper than the radio to optical spectrum).

At this time there are two major problems for the X-ray synchrotron
model based on a population of relativistic electrons described by a
single power law or a double power law (characterized by a `break
frequency').  The first of these arises from recent observations which
attempt to define the IR and optical spectra of individual features at
high spatial resolution.  For both Pictor A (the western hotspot) and
3C 273, cutoffs in the radio-optical spectra of the feature observed
in the X-rays demonstrate that a simple spectral extrapolation of the
\textit{spatial} emission at optical wavelengths cannot provide an
explanation for the observed X-ray intensity (R\"{o}ser et al. 1997).
Since straightforward SSC models also fail to predict the observed
X-rays, this means that we need a more complex spatial/spectral model
or we need a different emission process.  One could imagine very
compact components undetectable in the radio and/or optical because of
synchrotron self-absorption, but still producing X-ray synchrotron
emission.  Something along these lines would be consistent with the
variability results of Harris, Biretta, and Junor (1997) who predict
that the X-ray size of the M87 knot A will be substantially smaller
than that observed in the radio and optical.  Planned AXAF
observations will have the required resolution to test this
prediction.

\begin{figure}
\centerline{
\psfig{figure=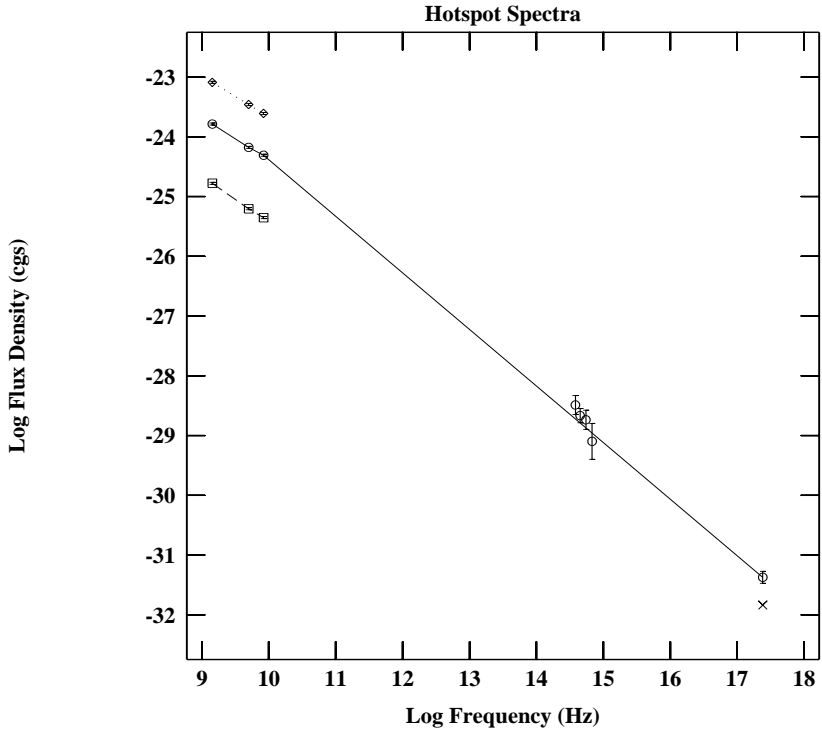,width=15cm,angle=-90}
}
\caption{\label{fig3} Hotspot spectra from radio to X-rays for 3C
390.3 (taken from Harris, Leighly, and Leahy, 1998).  The radio points
are VLA peak flux densities with a 2.8$^{\prime\prime}$ beamsize.  The
optical data are from Prieto and Kotilainen (1997).  The only X-ray
detection is for HS B; the `X' just below that point is an upper limit
for both the Np and Sf hotspots.  Circles are for HS B with a solid
line connecting the radio and X-ray points.  The dashed line between
the squares are for hotspot A (the Np hotspot) which, to avoid
confusion with HS B, are plotted a factor of 10 below the actual
values. The dotted line and diamonds are for the South following
hotspot. }
\end{figure}

The second problem for the standard (shock model) synchrotron spectrum
extending up to X-ray energies is to explain why some acceleration
sites are capable of producing enough electrons with energies
$\gamma~\approx~10^{7}$ to generate the observed X-ray emission,
whereas most hotspots and knots do not sustain these conditions.  One
common characteristic of these features appears to be the presence of
a very large gradient in the radio surface brightness, indicative of a
shock as the underlying cause of the radio feature.  For knot A in the
M87 jet, this leading edge feature is well known, both in the radio
and optical.  In figure~\ref{fig2}, we show another example, the
northern hotspot B in the FR II radio galaxy, 3C 390.3.  As discussed
in Harris, Leighly, and Leahy (1998), there is circumstantial evidence
that hotspot B owes its existence to the shock formed in the jet as it
enters the extended atmosphere of a dwarf galaxy.  New optical data
have been reported by Prieto \& Kotilainen (1997), and their images
demonstrate how well the shock front lies along the common edge of the
galaxy and the hotspot.  The currently available data are consistent
with a single power law from the radio to the X-ray
(figure~\ref{fig3}).  Thus a simple synchrotron model is acceptable,
with the electron spectrum extending up to $\gamma~\approx~7~10^{7}$
and an equipartition magnetic field strength of 44 $\mu$G leading to a
half-life of 57 years for the most energetic electrons.

\begin{figure}
\centerline{
\psfig{figure=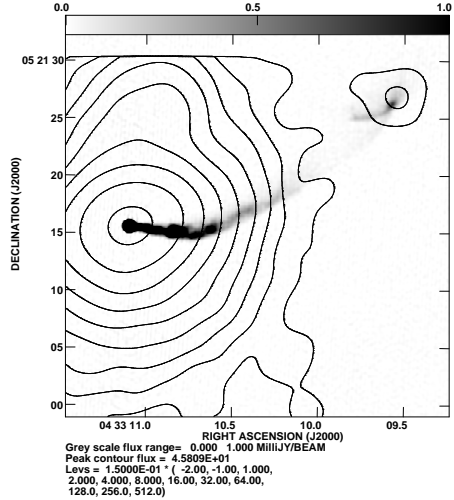,width=6cm,height=7.2cm}
}
\caption{\label{fig4} X-ray contours of 3C 120 from the ROSAT HRI
overlaid on a greyscale radio map from the VLA.  The X-ray map has
been smoothed with Gaussian of FWHM = 3$^{\prime\prime}$.  Contours are
logarithmic, increasing by factors of two.  The first contour level is
0.15 counts per 0.5$^{\prime\prime}$ pixel.  The radio map was kindly
provided by C. Walker. }
\end{figure}

Another example is a radio knot in the jet of 3C 120, shown in
Figure~\ref{fig4}. In this case, there is no optical emission
detected, nor is there any obvious reason for the existence of the
large gradient in radio surface brightness on the western edge of the
knot, yet it is this feature, not the gradient on the south edge,
which is associated with the X-ray emission.  Optical upper limits
preclude the construction of a radio to X-ray spectrum consisting of
one or two power laws.  Thus although we identify the X-ray emission
with the site of a shock, a simple synchrotron model is difficult to
construct.

\section{The Situation at High Redshift}

Although AGN X-ray emission can be detected at high redshift, the sort
of features described in the previous section are intrinsically too
weak for current detectors.  This is not true however for extended
thermal emission from clusters, and Carilli (this volume) describes a
ROSAT detection of a z=2.1 radio galaxy.  However, there were not
enough photons collected ($\approx$ 1 photon per ksec) to be sure
that the resulting spatial distribution is larger than the
instrumental point response function.  Whether or not well formed hot
cluster atmospheres will be found at large z remains to be determined.

The one condition that we may confidently expect to be different at
high redshift is the photon energy density of the cosmic background
which goes as (1+z)$^{4}$.  Since IC losses are proportional to the
energy density of photons, the half-life against IC losses at z=2.16
will be 100 times less than at z=0.  This effect may operate to limit
the physical sizes of radio sources at high redshift.  Krolik and Chen
(1991) have discussed the affect of redshift in that one samples
segments of the radio spectrum at higher and higher rest frame
frequencies as the redshift increases, thereby producing a statistical
increase in the mean spectral index.  However, the extremely steep
spectra ($\alpha\gg$1.2) found for some high z radio galaxies has not
been satisfactorily explained.  Perhaps these sources are heavily
weighted with components for which the exponential cutoff in the
electron spectrum is lower than for z=0 sources because of the much
larger E$^{2}$ IC/3k losses.

\section{Summary}

We reviewed the major ways that X-ray observations contribute to our
understanding of radio galaxies.  Many of these methods have been of
limited usefulness because current X-ray satellites have lacked the
combination of spatial and spectral resolution required to obtain
density and temperature information on a suitable scale to make
detailed comparison with the radio features.  In spite of this
problem, current X-ray data indicate that:

\begin{itemize}

\item	Most or all galaxies have a hot, extended atmosphere.

\item	Most or all radio features exclude ambient thermal plasma.

\item The most likely reasons that external thermal pressures are
	greater than internal minimum pressures are that the actual
	non-thermal pressures are substantially greater than the
	minimum values: significant contributions to the energy
	density come from protons and/or low energy electrons, neither
	of which can be directly measured.  Filling factors less than
	one are also probable in some sources.  We are not aware of
	any compelling evidence for a substantial departure from
	equipartition or for significant contributions to the internal
	pressure from thermal gas.

\item	With the above caveats, classical methods of estimating the
	equipartition magnetic field are probably not grossly in
	error.

\item	Buoyancy and large scale gas motions affect radio morphology,
	particularly for lower brightness features.

\item	Several signatures of the interaction of radio jets and lobes
	with the ambient gas have been found, and more should be
	forthcoming with the advent of AXAF.

\item	IC/3k emission from the radio lobes of Fornax A and SSC
	emission from the Cygnus A hotspots provide unique estimates
	of the average magnetic field strength.

\item	For some hotspots and knots in jets, simple synchrotron and
	inverse Compton models fail to provide for the observed X-ray
	intensity.

\item 	Unless characteristic magnetic field strengths are
	significantly larger in high redshift sources than they are
	locally, the (1+z)$^{4}$ increase in the photon energy density
	of the microwave background will increase the ratio of IC/3k
	(X-ray) to synchrotron (radio) power emitted by radio sources.

\end{itemize}
	
\begin{acknow}

We benefited from useful discussions with M. Birkinshaw, M. Elvis,
G. Fabbiano, W. Forman, C. Jones, D. Kim, and D. Worral.  The ROSAT
work on M87 comes from an ongoing collaboration with J. Biretta and
W. Junor; that on 3C 120 with J. Hjorth, A. Sadun, and M. Vestergaard.
We thank C. Carilli for helpful comments on the manuscript. NASA
contract NAS5-30934 and grant NAG5-2960 provided partial support.
\end{acknow}

\end{document}